\documentclass[a4paper,twoside,12pt]{article}
\input{obsjkt.sty}

\newcommand{\Msun}{\ensuremath{~{\rm M}_\odot}}                   
\newcommand{\Rsun}{\ensuremath{~{\rm R}_\odot}}                   
\newcommand{\rhosun}{\ensuremath{~\rho_\odot}}                    
\newcommand{\Teff}{\ensuremath{T_{\rm eff}}}                      
\newcommand{\Vsys}{\ensuremath{V_\gamma}}                         
\newcommand{\EBV}{\ensuremath{E(B\!-\!V)}}                        
\newcommand{\Grp}{\ensuremath{G_{\rm RP}}}                        
\newcommand{\degr}{\ensuremath{^\circ}}                           
\renewcommand{\kms}{~km~s$^{-1}$}                                 
\newcommand{\mc}[1]{\multicolumn{2}{c}{#1}}

\newcommand{\gaia}{\textit{Gaia}}
\newcommand{\targ}{HO~Tel}
\newcommand{\targfull}{HO~Telescopii}

\newcommand{\Msunnom}{\hbox{$\mathcal{M}^{\rm N}_\odot$}}
\newcommand{\Rsunnom}{\hbox{$\mathcal{R}^{\rm N}_\odot$}}
\newcommand{\Lsunnom}{\hbox{$\mathcal{L}^{\rm N}_\odot$}}

\usepackage{rotating}

\begin{document} 

\OBSheader{Rediscussion of eclipsing binaries: \targ}{J.\ Southworth}{2024 October}

\OBStitle{Rediscussion of eclipsing binaries. Paper XX. \\ HO Tel checkout}

\OBSauth{John Southworth}

\OBSinstone{Astrophysics Group, Keele University, Staffordshire, ST5 5BG, UK}


\OBSabstract{We present a detailed analysis of the detached eclipsing binary system \targfull, which contains two A-type stars in a circular orbit of period 1.613~d. We use light curves from the Transiting Exoplanet Survey Satellite (TESS), which observed \targ\ in three sectors, to determine its photometric properties and a precise orbital ephemeris. We augment these results with radial velocity measurements from S\"urgit et al.\ \cite{Surgit+17newa} to determine the masses and radii of the component stars: $M_{\rm A} = 1.906 \pm 0.031$\Msun, $M_{\rm B} = 1.751 \pm 0.034$\Msun, $R_{\rm A} = 2.296 \pm 0.027$\Rsun\ and $R_{\rm B} = 2.074 \pm 0.028$\Rsun. Combined with temperature measurements from S\"urgit et al.\ \cite{Surgit+17newa} and optical-infrared apparent magnitudes from the literature, we find a distance to the system of $280.8 \pm 4.6$~pc which agrees well with the distance from the \gaia\ DR3 parallax measurement. Theoretical predictions do not quite match the properties of the system, and there are small discrepancies in measurements of the spectroscopic orbits of the stars. Future observations from \gaia\ will allow further investigation of these issues.}


\section*{Introduction}

In the current series of papers we are performing detailed photometric analyses of a set of known detached eclipsing binaries (dEBs) for which space-based light curves are available but have not been studied previously, and which have published spectroscopic mass measurements. The aim is to increase the number of stars with precisely-measured masses and radii against which theoretical stellar models can be compared \cite{Andersen91aarv,Torres++10aarv,LastennetVallsgabaud02aa,DelburgoAllende18mn}. A detailed exposition of these goals can be found in the first paper of the series (ref.\ \citenum{Me20obs}) and a review of the impact of space telescopes in this scientific area can be found in ref.~\citenum{Me21univ}.

In this work we investigate the dEB \targfull\ (Table~\ref{tab:info}), which contains two late-A type stars in a circular orbit of period 1.613~d. Its variability was discovered by Strohmeier et al.\ \cite{Strohmeier++65ibvs} under the designation BV~590, and its correct orbital period was determined by Spoelstra \& van Houten \cite{SpoelstraVanhouten72aas}. Subsequent work on this object has been nicely summarised by S\"urgit et al.\ \cite{Surgit+17newa} (hereafter S17). These authors presented radial velocity (RV) measurements from medium-resolution spectra obtained with the SpUpNIC spectrograph \cite{Crause+19jatis} on the 74-inch Radcliffe Telescope at the South African Astronomical Observatory. S17 combined these RVs with five-colour (Walraven \cite{WalravenWalraven60ban} $VBLUW$) light curves from Spoelstra \& van Houten \cite{SpoelstraVanhouten72aas} and the light curve from the All-Sky Automated Survey (ASAS \cite{Pojmanski97aca}) to measure the properties of the system. Below we use the same RVs and new space-based data to refine the measurements of the system properties.



\begin{table}[t]
\caption{\em Basic information on \targfull. The $BV$ magnitudes are each the mean of 110 individual measurements \cite{Hog+00aa}. \label{tab:info}}
\centering
\begin{tabular}{lll}
{\em Property}                            & {\em Value}                 & {\em Reference}                      \\[3pt]
Right ascension (J2000)                   & 19 51 58.93                 & \citenum{Gaia21aa}                   \\
Declination (J2000)                       & $-$46 51 42.4               & \citenum{Gaia21aa}                   \\
Henry Draper designation                  & HD 187418                   & \citenum{CannonPickering23anhar}     \\
\textit{Gaia} DR3 designation             & 6671501451113955072         & \citenum{Gaia21aa}                   \\
\textit{Gaia} DR3 parallax                & $3.5186 \pm 0.0314$ mas     & \citenum{Gaia21aa}                   \\          
TESS\ Input Catalog designation           & TIC 80064289                & \citenum{Stassun+19aj}               \\
$B$ magnitude                             & $8.59 \pm 0.03$             & \citenum{Hog+00aa}                   \\          
$V$ magnitude                             & $8.31 \pm 0.01$             & \citenum{Hog+00aa}                   \\          
$J$ magnitude                             & $7.814 \pm 0.027$           & \citenum{Cutri+03book}               \\
$H$ magnitude                             & $7.776 \pm 0.029$           & \citenum{Cutri+03book}               \\
$K_s$ magnitude                           & $7.730 \pm 0.018$           & \citenum{Cutri+03book}               \\
Spectral type                             & A7~V + A8~V                 & \citenum{Surgit+17newa}              \\[3pt]
\end{tabular}
\end{table}


\section*{Photometric observations}

\begin{figure}[t] \centering \includegraphics[width=\textwidth]{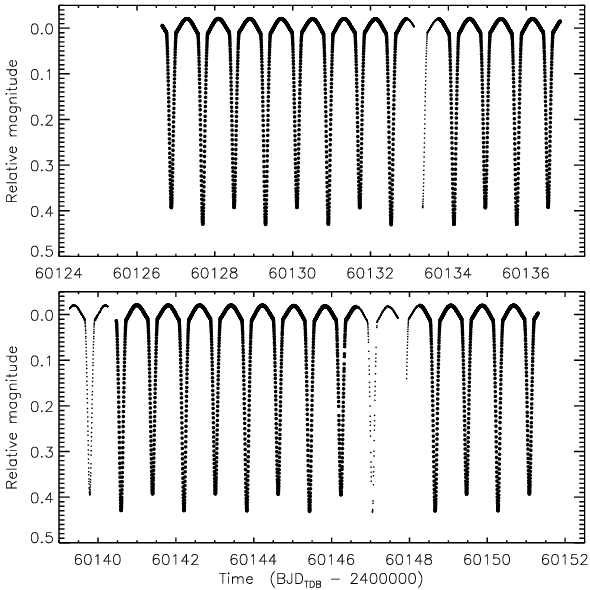} \\
\caption{\label{fig:time} TESS short-cadence SAP photometry of \targ\ from sector 67. 
The flux measurements have been converted to magnitude units then rectified to zero 
magnitude by subtraction of the median. The two panels show half the sector each.
Larger points show the data retained for analysis and smaller points the data
rejected due to offsets or increased scatter.} \end{figure}

%

\targ\ has been observed in three sectors by the Transiting Exoplanet Survey Satellite (TESS \cite{Ricker+15jatis}): sector 13 (2019 July) where the observations were summed into cadences of 1800~s duration; sector 27 with a cadence of 600~s; and sector 67 with a cadence of 200~s. We downloaded the data for all sectors from the NASA Mikulski Archive for Space Telescopes (MAST\footnote{\texttt{https://mast.stsci.edu/portal/Mashup/Clients/Mast/Portal.html}}) using the {\sc lightkurve} package \cite{Lightkurve18}. However, we restricted our analysis below to the data from sector 67 due to its better sampling rate. We adopted the simple aperture photometry (SAP) data from the TESS-SPOC data reduction \cite{Jenkins+16spie} with a quality flag of ``hard''. These were normalised using {\sc lightkurve} and converted to differential magnitude.

The light curve from sector 67 is shown in Fig.~\ref{fig:time}. Four regions of data (one of which is outside the figure) were removed from our analysis due to incomplete coverage of eclipses or decreased photometric precision due to scattered light from Earth: we kept 8006 of the original 9332 datapoints. The primary eclipse is clearly deeper than the secondary. We label the star eclipsed during primary minimum star~A and its companion star~B.

We queried the \gaia\ DR3 database\footnote{\texttt{https://vizier.cds.unistra.fr/viz-bin/VizieR-3?-source=I/355/gaiadr3}} and found a total of 75 objects within 2~arcmin of the sky position of \targ. Of these, the brightest is fainter than our target by 4.5~mag in the \Grp\ band, and the remainder are fainter by at least 5.7~mag in this band. This suggests that the TESS light curve of \targ\ will be contaminated by light from nearby stars at the level of only a few percent.


\section*{Preliminary light curve analysis}

The components of \targ\ are close and significantly distorted from sphericity. However, the number of datapoints is large enough to make an analysis with a code implementing Roche geometry slow. We have therefore undertaken a preliminary analysis with a simpler code to determine the orbital ephemeris and enable the construction of a phase-binned light curve.

We modelled the TESS sector 67 light curve using version 43 of the {\sc jktebop}\footnote{\texttt{http://www.astro.keele.ac.uk/jkt/codes/jktebop.html}} code \cite{Me++04mn2,Me13aa} using a suitable set of adjustable parameters (see previous papers in this series). Once a good fit was achieved, the TESS observations were converted to orbital phase and binned into 1000 points equally-spaced in phase. This phase-binned light curve retains practically all the information of the original data whilst containing a factor of eight fewer datapoints.


\section*{Orbital ephemeris}

We refined the orbital ephemeris of \targ\ by adding new and published times of mid-eclipse to our {\sc jktebop} fit. We included the four times from Sistero \& Candellero \cite{SisteroCandellero79ibvs}, and the four times from Spoelstra \& Van Houten \cite{SpoelstraVanhouten72aas}. Uncertainties were not quoted for these measurements so we adopted an errorbar of $\pm$0.003~d for each. We also measured three additional times of primary eclipse by fitting the TESS sectors individually. The precision of these eclipse times is extraordinary (0.3 to 0.9~s) but appears to be justified. The early times were converted to the BJD$_{\rm TDB}$ timescale \cite{Eastman++10pasp} to match the TESS data.

We also tried to include the timing from table~2 of S17 but found it to deviate from a linear ephemeris by +30.8~min; conversion from the original HJD (presumed UTC) to the BJD$_{\rm TDB}$ timescale used in the current paper would add a further 65~s to this discrepancy. The issue probably arises from the use of an old time of conjunction combined with a fixed period, which is not a problem for fitting the RV cuve but does make the timing unsuitable for determining the orbital period. We therefore discluded it from our analysis.

The ephemeris was obtained as part of our {\sc jktebop} solution in the preceding section and is
\begin{equation}
{\rm Min~I} ~=~ {\rm BJD}_{\rm TDB}\,2460135.755972 (3) + E\times1.613103937 (8)
\end{equation}
where $E$ is the cycle number and the bracketed quantities represent the uncertainty in the final digit of the preceding number. The individual eclipse times and their residuals versus this linear ephemeris are given in Table~\ref{tab:tmin}. We see no evidence in these data for a deviation from a constant orbital period.

\begin{table} \centering
\caption{\em Times of published mid-eclipse for \targ\ and their residuals versus the best fit reported
in the current work. Each residual is given as a fraction of the uncertainty. The asterisk indicates the 
time not included in the final best fit, to avoid double-use of data. The orbital cycle is an integer 
for primary eclipses and a half-integer for secondary eclipses. \label{tab:tmin}}
\begin{tabular}{lllllll}
{\em Orbital}        & {\em Eclipse time} & {\em Uncertainty} & {\em Best fit}    & Residual      & {\em Source} \\
{\em cycle}          & (BJD$_{\rm TDB}$)  & {\em (d)}         & (BJD$_{\rm TDB}$) & ~~~($\sigma$) &              \\[3pt]
$-$13113.5           & 2438982.31756  & 0.003    & 2438982.31750  & \phantom{$-$}0.12 & \cite{SisteroCandellero79ibvs} \\
$-$13111             & 2438986.34984  & 0.003    & 2438986.35026  & \phantom{$-$}0.28 & \cite{SisteroCandellero79ibvs} \\
$-$13108.5           & 2438990.38244  & 0.003    & 2438990.38302  & \phantom{$-$}0.34 & \cite{SisteroCandellero79ibvs} \\
$-$13087.5           & 2439024.25888  & 0.003    & 2439024.25821  &           $-$0.08 & \cite{SisteroCandellero79ibvs} \\
$-$11072.5           & 2442274.65961  & 0.003    & 2442274.66264  & \phantom{$-$}1.18 & \cite{SpoelstraVanhouten72aas} \\
$-$11072             & 2442275.47641  & 0.003    & 2442275.46919  &           $-$2.24 & \cite{SpoelstraVanhouten72aas} \\
$-$10886.5           & 2442574.70254  & 0.003    & 2442574.69997  &           $-$0.68 & \cite{SpoelstraVanhouten72aas} \\
$-$10649             & 2442957.81126  & 0.003    & 2442957.81216  & \phantom{$-$}0.48 & \cite{SpoelstraVanhouten72aas} \\
  $-$909             & 2458669.444500 & 0.000010 & 2458669.444494 &           $-$0.61 & This work \\
  $-$673             & 2459050.137019 & 0.000007 & 2459050.137023 & \phantom{$-$}0.57 & This work \\
\phantom{$-$}0\,$^*$ & 2460135.755971 & 0.000003 &                &                   & This work \\
\end{tabular}
\end{table}


\section*{Analysis with the Wilson-Devinney code}

The main analysis of the light curve was performed using the Wilson-Devinney (WD) code \cite{WilsonDevinney71apj,Wilson79apj}, which implements Roche geometry to accurately represent the shapes of distorted stars. We used the 2004 version of the code ({\sc wd2004}), driven by the {\sc jktwd} wrapper \cite{Me+11mn}, to fit the phase-binned light curve from the previous section. Below we describe the adopted solution of the light curve, followed by the error analysis. The parameters in the {\sc wd2004} code are described in its accompanying user manual (ref.~\cite{WilsonVanhamme04}).

\begin{figure}[t] \centering \includegraphics[width=\textwidth]{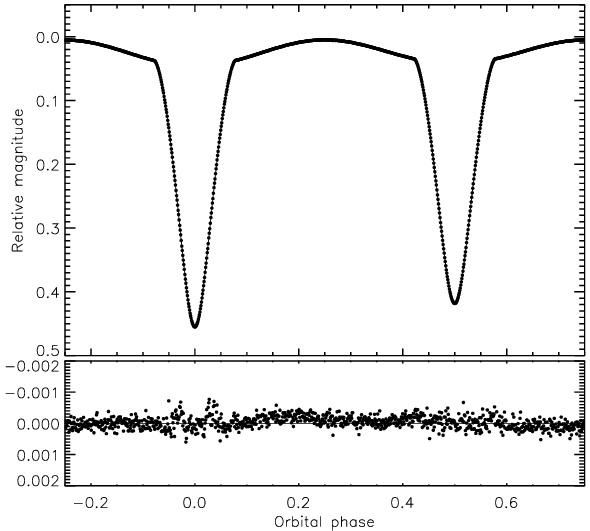} \\
\caption{\label{fig:phase} Best fit to the binned light curve of \targ\ using {\sc wd2004}.
The phase-binned data are shown using open circles and the best fit with a continuous line.
The residuals are shown on an enlarged scale in the lower panel.} \end{figure}

For our adopted solution we fitted for the potentials and light contributions of the two stars, the orbital inclination, and one limb darkening coefficient per stars. Limb darkening was implemented using the logarithmic law with the linear coefficients fitted and the nonlinear coefficients fixed at theoretical values from Van Hamme \cite{Vanhamme93aj}. We also had to fit for the albedo of both stars and for third light to obtain a good fit to the data. We used mode 0, where the effective temperatures (\Teff s) and light contributions are decoupled, and fixed the \Teff s to values from S17. We adopted a circular orbit, the mass ratio from S17, the simple model of reflection, synchronous rotation, gravity darkening exponents of 1.0 (suitable for radiative admospheres), the maximum possible numerical precision of {\sc n1} $=$ {\sc n2} $=$ 60, and the Johnson $R$ passband as representative of the TESS passband for stars like those in \targ. With this approach we obtained a good fit to the data (Fig.~\ref{fig:phase}) which has residuals that are small but do show a trend with orbital phase. The parameters of this fit are given in Table~\ref{tab:wd}.

\begin{table} \centering
\caption{\em Summary of the parameters for the {\sc wd2004} solution of the TESS light curve of \targ. Uncertainties
are only quoted when they have been assessed by comparison between a full set of alternative solutions. \label{tab:wd}}
\begin{tabular}{lcc}
{\em Parameter}                           & {\em Star A}          & {\em Star B}          \\[3pt]           
{\it Control parameters:} \\
{\sc wd2004} operation mode               & \multicolumn{2}{c}{0}                         \\                
Treatment of reflection                   & \multicolumn{2}{c}{1}                         \\                
Number of reflections                     & \multicolumn{2}{c}{1}                         \\                
Limb darkening law                        & \multicolumn{2}{c}{2 (logarithmic)}           \\                
Numerical grid size (normal)              & \multicolumn{2}{c}{60}                        \\                
Numerical grid size (coarse)              & \multicolumn{2}{c}{60}                        \\[3pt]           
{\it Fixed parameters:} \\
Phase shift                               & \multicolumn{2}{c}{0.0}                       \\                
Mass ratio                                & \multicolumn{2}{c}{0.921}                     \\                
Rotation rates                            & 1.0                   & 1.0                   \\                
Gravity darkening                         & 1.0                   & 1.0                   \\                
\Teff\ values (K)                         & 7872                  & 7627                  \\                
Bolometric linear LD coefficient          & 0.6720                & 0.6799                \\                
Bolometric logarithmic LD coefficient     & 0.1991                & 0.2043                \\                
Passband logarithmic LD coefficient       & 0.2454                & 0.2430                \\[3pt]           
{\it Fitted parameters:} \\
Bolometric albedos                        & $1.20 \pm 0.40$       & $1.00 \pm 0.24$       \\                
Potential                                 & $4.848 \pm 0.048$     & $5.036 \pm 0.051$     \\                
Orbital inclination (\degr)               & \multicolumn{2}{c}{$81.041 \pm 0.067$}        \\                
Light contributions                       & $6.92 \pm 0.19$       & $5.19 \pm 0.18$       \\                
Passband linear LD coefficient            & $0.579 \pm 0.017$     & $0.550 \pm 0.017$     \\                
Third light                               & \multicolumn{2}{c}{$0.016 \pm 0.013$}         \\                
{\it Derived parameters:} \\
Fractional radii                          & $0.2575 \pm 0.0026$   & $0.2325 \pm 0.0029$   \\                
Light ratio                               & \multicolumn{2}{c}{$0.754 \pm 0.036$}         \\[10pt]          
\end{tabular}
\end{table}

The errorbars returned by {\sc wd2004} account for the scatter of the data but not for the many choices made during the modelling process, so are far too small. To determine realistic errorbars we performed a large number of alternative modelling runs whilst varying the input physics and treatment of the data. These differences were: using mode 2 and fitting for \Teff\ instead of the light contribution of star~B; changing the spectroscopic mass ratio by its uncertainty; changing the rotation rates by $\pm$0.1, changing the gravity darkening exponents by $\pm$0.1; fitting for a phase shift; fixing the limb darkening coefficients at the theoretically-predicted values; using the square-root limb darkening law; using the Johnson $I$ passband instead of $R$; changing the numerical precision values ({\sc n1} and {\sc n2}) to 59, 58, 57 or 56; using the detailed reflection effect option; using two instead of one reflection with the detailed reflection treatment; using a light curve phase-binned into 500 instead of 1000 points; and removing the polynomials from the {\sc jktebop} fit before binning. This process is basically the same as has been used for multiple systems in the past \cite{Me+20mn,MeBowman22mn,Me22obs4,Jennings+24mn}.

The result of this process was a large set of different parameter values. The differences for each parameter versus the adopted solution were added in quadrature to obtain the final uncertainty for that parameter. These errorbars are reported in Table~\ref{tab:wd}. The albedos and third light values are quite uncertain: their errorbars are dominated by the variation obtained when using the Johnson $I$ band instead of the $R$ band. 

The fractional radii of the stars are determined to 1.0\% and 1.3\% precision, respectively, but the main source of uncertainty is unexpected. To illustrate this we give in Table~\ref{tab:rerror} the individual contributions to the uncertainties in the fractional radii which arise from the various model choices listed above. The largest effect is due to the choice of numerical precision, which sets a limit on how well the fractional radii ($r_{\rm A}$ and $r_{\rm B}$) can be measured. We have previously seen this effect in our analysis of the eclipsing system KIC~4851217 (paper in preparation) so the current result is not an isolated incident. It is likely that more sophisticated modelling codes \cite{Prsa+16apjs} will suffer less from this effect and thus allow an increase in the precision achievable in the determination of the properties of distorted stars in eclipsing binary systems.

\begin{table} \centering
\caption{\em Changes in the measured fractional radii of the stars due to differing model choices. 
Each is expressed as the percentage change versus the value of the parameter. \label{tab:rerror}}
\begin{tabular}{lrr}
Model choice                                      &       \mc{~Effect (\%)}     \\
                                                  & $r_{\rm A}$~ &  $r_{\rm B}$~ \\[3pt]
Changing mass ratio                                         &   0.38 & $-$0.56 \\
Changing rotation rates by $\pm$0.1                         &   0.28 & $-$0.21 \\
Changing gravity darkening by $\pm$0.1                      &   0.03 & $-$0.02 \\
Fitting for phase shift                                     &   0.00 &    0.00 \\
Fixing limb darkening coefficients                          &   0.34 & $-$0.18 \\
Using the square-root limb darkening law                    &$-$0.01 & $-$0.04 \\
Using the Johnson $I$-band                                  &   0.10 & $-$0.16 \\
Setting the numerical precision to {\sc n1}$=${\sc n2}$=$58 &   0.82 & $-$1.02 \\
Using the detailed treatment of reflection                  &   0.01 &    0.00 \\
Detailed treatment of reflection with two reflections       &   0.01 & $-$0.00 \\
Modelling a light curve of 500 phase-binned datapoints      &   0.09 & $-$0.13 \\
Removing the polynomial normalisation                       &   0.05 & $-$0.19 \\
\end{tabular}
\end{table}


\section*{Radial velocity analysis}

S17 obtained 45 medium-resolution spectra, from each of which they measured RVs for both stars using cross-correlation. These were included in their analysis with the WD code and the resulting parameters were given with 90\% confidence intervals. As we universally use standard errors we have reanalysed the RVs to determine our own spectroscopic orbital parameters.

\begin{figure}[t] \centering \includegraphics[width=\textwidth]{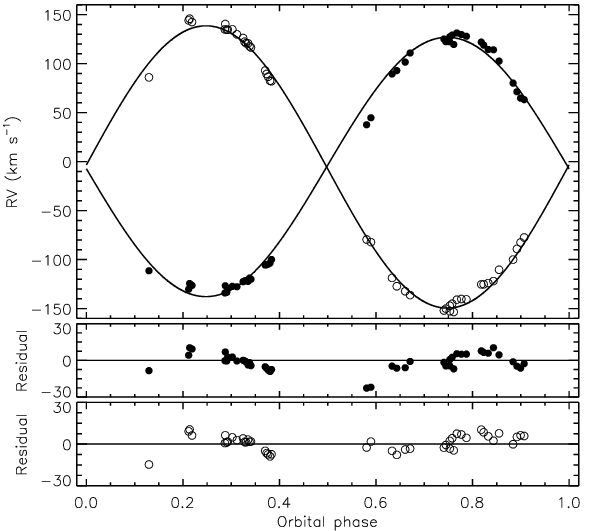} \\
\caption{\label{fig:rv} RVs of \targ\ from S17 (filled circles for star~A and open 
circles for star~B), compared to the best fit from {\sc jktebop} (solid lines). The 
residuals are given in the lower panels separately for the two components.} \end{figure}

The RVs were obtained from table~1 in S17 and modelled using the {\sc jktebop} code, with the orbital ephemeris from above but with no other constraints from the TESS light curve. The quoted errorbars were scaled so the fit to the RVs of each star had a reduced $\chi^2$ value of unity. We fitted for the velocity amplitudes of the stars and the systemic velocity of the system, obtaining $K_{\rm A} = 132.3 \pm 1.2$\kms, $K_{\rm B} = 144.0 \pm 1.0$\kms\ and $\Vsys = -5.5 \pm 0.6$\kms, respectively. The errorbars for these quantities were obtained using Monte Carlo simulations \cite{Me21obs5}. If the systemic velocities of the two stars are fitted separately there is a difference of 3.9\kms\ between the stars, and the $K_{\rm A}$ and $K_{\rm B}$ change by $-$0.4\kms.

The best fit to the RVs is shown in Fig.~\ref{fig:rv}. It can be seen that there are three spectra which give large residuals, around phases 0.13 and 0.59. If these are rejected the measured properties become $K_{\rm A} = 133.9 \pm 1.2$\kms, $K_{\rm B} = 145.5 \pm 0.8$\kms\ and $\Vsys = -4.4 \pm 0.5$\kms. We chose not to adopt these values, because there were no clear reasons to reject those data, but report them for completeness.

The spectroscopic orbital parameters given by S17 agree with our own results, and are $K_{\rm A} = 131.5 \pm 1.2$\kms, $K_{\rm B} = 142.8 \pm 1.2$\kms\ and $\Vsys = -5.9 \pm 0.7$\kms. A cross-check of these numbers is also available using the \gaia\ \cite{Gaia16aa} DR3 \cite{Gaia23aa} \texttt{tbosb2} catalogue\footnote{\texttt{https://vizier.cds.unistra.fr/viz-bin/VizieR-3?-source=I/357/tbosb2}} \cite{Gaia23aa2}, which includes the parameters of a double-lined spectroscopic orbit for the system. Based on 12 RVs for each star the orbit is $K_{\rm A} = 136.8 \pm 1.2$\kms, $K_{\rm B} = 144.5 \pm 1.2$\kms\ and $\Vsys = -7.5 \pm 0.6$\kms. The velocity amplitude of star~A is somewhat higher than that found from the RVs of S17, but this cannot be investigated further because the individual RVs from \gaia\ have not be made publicly available. A small disagreement was also found in our analysis of V570~Per \cite{Me23obs4}, and other issues have been noted in the literature \cite{Bashi+22mn,Tokovinin23aj,MarcussenAlbrecht23aj}, so we look forward to the RV measurements and individual spectra becoming available in future.


\section*{Physical properties and distance to \targ}

\begin{table} \centering
\caption{\em Physical properties of \targ\ defined using the nominal solar units given by 
IAU 2015 Resolution B3 (ref.\ \cite{Prsa+16aj}). \label{tab:absdim}}
\begin{tabular}{lr@{\,$\pm$\,}lr@{\,$\pm$\,}l}
{\em Parameter}        & \multicolumn{2}{c}{\em Star A} & \multicolumn{2}{c}{\em Star B}    \\[3pt]
Mass ratio   $M_{\rm B}/M_{\rm A}$          & \multicolumn{4}{c}{$0.919 \pm 0.010$}         \\
Semimajor axis of relative orbit (\Rsunnom) & \multicolumn{4}{c}{$8.918 \pm 0.050$}         \\
Mass (\Msunnom)                             &  1.906  & 0.031       &  1.751  & 0.034       \\
Radius (\Rsunnom)                           &  2.296  & 0.027       &  2.074  & 0.028       \\
Surface gravity ($\log$[cgs])               &  3.996  & 0.009       &  4.048  & 0.012       \\
Density ($\!\!$\rhosun)                     &  0.1574 & 0.0048      &  0.1965 & 0.0075      \\
Synchronous rotational velocity ($\!\!$\kms)& 72.03   & 0.83        & 65.03   & 0.89        \\
Effective temperature (K)                   &  7872   & 200         &  7627   & 201         \\
Luminosity $\log(L/\Lsunnom)$               &   1.261 & 0.045       &  1.118  & 0.047       \\
$M_{\rm bol}$ (mag)                         &   1.59  & 0.11        &  1.95   & 0.12        \\
Interstellar reddening \EBV\ (mag)			& \multicolumn{4}{c}{$0.04 \pm 0.02$}			\\
Distance (pc)                               & \multicolumn{4}{c}{$280.8 \pm 4.6$}           \\[3pt]
\end{tabular}
\end{table}


The physical properties of \targ\ were determined from the results of the {\sc wd2004} code and RV analyses given above, using the {\sc jktabsdim} code \cite{Me++05aa} (Table~\ref{tab:absdim}). The masses and radii of the component stars are now known to 2\% or better, matching the minimum requirements for a useful comparison with theoretical models \cite{Andersen91aarv,Me15aspc}. Our results agree well with those from S17, but the availability of the TESS data has allowed us to improve the measurement precision of the radii from 7\% to 1\%.

To determine the distance to the system we adopted the \Teff\ measurements from S17, the $BV$ and $JHK_s$ magnitudes from Table~\ref{tab:info}, the surface brightness calibrations from Kervella et al.\ \cite{Kervella+04aa} and the method from Southworth et al.\ \cite{Me++05aa}. The 2MASS $JHK_s$ magnitudes were obtained at orbital phase 0.796. A small interstellar reddening of $\EBV = 0.04 \pm 0.02$ was needed to bring the distances from the $BV$ bands into agreement with those from the $JHK_s$ bands. The resulting distance of $280.8 \pm 4.6$~pc agrees with the value of $284.2 \pm 3.2$~pc from the \gaia\ DR3 parallax.


\section*{Conclusion}

The dEB \targ\ contains two A-type stars in a short-period orbit which causes them to be tidally deformed. We have determined their masses and radii using photometry from the TESS mission and published ground-based RVs from S17. The measurements are to 1.6\% and 1.9\% precision in mass, and 1.2\% and 1.0\% precision in radius. The mass measurements are limited by the scatter in the available RVs, and the radius measurements by the numerical precision of the modelling code used. Adding published \Teff s and apparent magnitudes to the analysis allowed a measurement of the distance to the system of $281 \pm 5$~pc, in agreement with the distance from \gaia\ DR3. 

We have compared the measured properties of the component stars to the predictions of the {\sc parsec} theoretical stellar evolutionary models \cite{Bressan+12mn}. We confirm the discrepancy found by S17 in that a good fit to both stars cannot be obtained for a single age, and that star~B matches predictions for older ages ($1030 \pm 70$~Myr) than star~A ($880 \pm 60$~Myr) for a solar metal abundance ($Z = 0.017$). The improved agreement seen by S17 in the Hertzsprung-Russell diagram suggests the discrepancy is related to the measured masses of the stars. To test this we used the $K_{\rm A}$ and $K_{\rm B}$ values from \gaia\ to obtain slightly higher masses and a good fit in the mass--radius diagram for an age of $800 \pm 50$~Myr. However, this results in an increase in the predicted \Teff s, which must then be brought down by adopting a higher metallicity of at least $Z=0.03$. 

\targ\ would benefit from more detailed spectroscopic study. Forthcoming data releases from \gaia\ will contain more epochs of spectroscopy, and the individual RV measurements from the RVS spectrometer, so will help in this work. Ground-based spectra would also be useful in determining the \Teff s and photospheric chemical compositions of the stars to better precision and accuracy.


\section*{Acknowledgements}

We thank the anonymous referee for a positive and extraordinarily prompt report.
This paper includes data collected by the TESS\ mission and obtained from the MAST data archive at the Space Telescope Science Institute (STScI). Funding for the TESS\ mission is provided by the NASA's Science Mission Directorate. STScI is operated by the Association of Universities for Research in Astronomy, Inc., under NASA contract NAS 5–26555.
This work has made use of data from the European Space Agency (ESA) mission {\it Gaia}\footnote{\texttt{https://www.cosmos.esa.int/gaia}}, processed by the {\it Gaia} Data Processing and Analysis Consortium (DPAC\footnote{\texttt{https://www.cosmos.esa.int/web/gaia/dpac/consortium}}). Funding for the DPAC has been provided by national institutions, in particular the institutions participating in the {\it Gaia} Multilateral Agreement.
The following resources were used in the course of this work: the NASA Astrophysics Data System; the SIMBAD database operated at CDS, Strasbourg, France; and the ar$\chi$iv scientific paper preprint service operated by Cornell University.


\bibliographystyle{obsmaga}
\bibliography{jkt}

\end{document}